\documentclass[manuscript,screen]{acmart}

\DeclareUnicodeCharacter{2060}{\nolinebreak}
\AtBeginDocument{%

  \providecommand\BibTeX{{%

\normalfont B\kern-0.5em{\scshape i\kern-0.25em b}\kern-0.8em\TeX}}}

\setcopyright{rightsretained}

\copyrightyear{2024}

\acmYear{2024}

\acmDOI{}

\acmConference[CCAI 2024]{CHI 2024 Workshop on Child-centred AI Design}{May 11, 2024}{Honolulu, HI, USA}

\acmBooktitle{CHI 2024 Workshop on Child-centred AI Design, May 11, 2024, Honolulu, HI, USA}

\begin{document}

\title{Toward Safe Evolution of Artificial Intelligence (AI) based Conversational Agents to Support Adolescent Mental and Sexual Health Knowledge Discovery}


\author{Jinkyung Park}
\affiliation{%
  \institution{Vanderbilt University}
  \city{Nashville}
  \country{USA}}
\email{jinkyung.park@vanderbilt.edu}

\author{Vivek Singh}
\affiliation{%
  \institution{Rutgers University}
  \city{New Brunswick}
  \country{USA}}
\email{v.singh@rutgers.edu}

\author{Pamela Wisniewski}
\affiliation{%
  \institution{Vanderbilt University}
  \city{Nashville}
  \country{USA}}
\email{pamela.wisniewski@vanderbilt.edu}
\renewcommand{\shortauthors}{Park et al.}
\renewcommand{\shorttitle}{Adolescent Mental and Sexual Health Knowledge Discovery through AI-based CAs}
\begin{abstract}
  Following the recent release of various Artificial Intelligence (AI) based Conversation Agents (CAs), adolescents are increasingly using CAs for interactive knowledge discovery on sensitive topics, including mental and sexual health topics. Exploring such sensitive topics through online search has been an essential part of adolescent development, and CAs can support their knowledge discovery on such topics through human-like dialogues. Yet, unintended risks have been documented with adolescents' interactions with AI-based CAs, such as being exposed to inappropriate content, false information, and/or being given advice that is detrimental to their mental and physical well-being (e.g., to self-harm). In this position paper, we discuss the current landscape and opportunities for CAs to support adolescents' mental and sexual health knowledge discovery. We also discuss some of the challenges related to ensuring the safety of adolescents when interacting with CAs regarding sexual and mental health topics. We call for a discourse on how to set guardrails for the safe evolution of AI-based CAs for adolescents. 
\end{abstract}




\maketitle

\section{Introduction}
Exploring sensitive topics such as sexual health topics through online search has been an essential part of adolescent (ages 13-17) development~\cite{razi2020let, gardner2013app}. 
Research has been conducted to understand adolescents' search behaviors to design safer web search tools~\cite{foss2014children}. These studies highlighted the importance of the social aspects of adolescents' web searches and called for search systems that can support them with diverse technical skills, reading levels, domain knowledge, and personal interests. Yet, the limitations of traditional safe search approaches include a heavy reliance on keyword-based filtering and a lack of flexibility to address the evolving online landscape~\cite{medium2023}. 
Following the recent release of various conversation agents (e.g.,  Microsoft Bing~\cite{Bing2023}, Google Bard ~\cite{Bard2023}, and OpenAI GPT-4~\cite{OpenAI2023}), online search has transformed into a more interactive process where users can engage in back-and-forth conversations with AI-based systems to discover knowledge~\cite{GoogleSearch2023, preininger2021differences, garg2022last}. 

Conversational Agents (CAs, often called chatbots) are systems enabled with the ability to interact with the users using natural human dialogue ~\cite{rheu2021systematic}. Now, CAs are increasingly used by adolescents for interactive knowledge discovery on sensitive topics, including mental and sexual health topics~\cite{sanabria2023great}. 
This shift underscores a significant gap in traditional safe web search research, as approaches for safe search no longer apply or cater to the nuanced demands of today's conversational AI systems. Consequently, there is a pressing need for new strategies to ensure safe interactive knowledge discovery within the context of conversational agents~\cite{Whitehouse2023}, addressing the unique challenges posed by this evolving technology. Recently, never-before encountered risks have been documented with teens’ interactions with AI-based CAs, such as being exposed to sexually inappropriate content and/or being given advice that is detrimental to their mental and physical wellbeing (e.g., to self-harm) ~\cite{Kelly2023, Fowler2023}. 

Considering the existing and potential harms that CAs can pose to adolescents, we need to start thinking about how to build guardrails to keep CAs safe for adolescents, especially related to sensitive topics, such as mental and sexual health. 
In this position paper, we present the current landscape of mental and sexual health CAs designed for adolescents, their potential benefits and challenges, and call for further research.  Our position paper is highly relevant to the Child-Centred AI (CCAI) workshop as our focus on safety in AI-based CAs for adolescents is one of the core topics of interest for CCAI. (i.e., predominant challenges and practical safeguards for translating child-centered AI concepts into practice). 

\section{Current Landscape}


\paragraph{\textbf{Conversational Agents to Support Adolescent Mental Health}}
As adolescents do not seek professional help for sensitive health problems due to reasons such as perceived social stigma and embarrassment, confidentiality, and financial costs~\cite{radez2021children}, CAs are increasingly applied to support the mental health and well-being of adolescents in a variety of contexts such as depression~\cite{kuhlmeier2022personalized}, anxiety~\cite{gabrielli2021engagement}, stress~\cite{williams202121}, and overall mental well-bing~\cite{maenhout2021participatory}. A plethora of work has been done to explore the feasibility and/or effectiveness of CAs to improve mental health conditions~\cite{dosovitsky2023development, kuhlmeier2022personalized, elmasri2016conversational, abreu2022raising}, educate how to promote mental well-being~\cite{koulouri2022chatbots, nicol2022chatbot, grove2021co, gabrielli2020chatbot}, and provide credible information/resources related to mental health~\cite{sanabria2023great, lescano2022iterative, beilharz2021development}. 
The major benefits of such mental health CAs that were documented from the early studies include serving as accessible alternatives for adolescents who are not comfortable with in-person conversations about their mental health needs. Adolescents in such situations benefit from access to informational resources~\cite{sanabria2023great} and emotional support by friendly and empathic, yet knowledgeable responses generated by CAs ~\cite{hoiland2020hi}.  In addition, therapeutic content based on Cognitive Behavioral Therapy (CBT) and positive psychology provided by mental health CAs has been shown to be effective in emotional relief for adolescents~\cite{he2022mental, nicol2022chatbot}. 

\paragraph{\textbf{Conversational Agents to Support Adolescent Sexual Health}}

While CAs can be used for sexual health information seeking especially at-risk adolescent populations such as sexual minority adolescents, who are even less likely to seek professional care due to limited resources and social stigmatization~\cite{balaji2022effectiveness, harb2019motivators}, compared to mental health, CAs to support adolescent sexual health information seeking are under-studied.  
Recently, a few sexual health CAs for adolescents have been studied to facilitate adolescents' sexual knowledge discovery~\cite{nadarzynski2021barriers, lescano2022iterative, rahman2021adolescentbot, wang2022artificial, bonnevie2021layla, massa2023transgender}. Most of the sexual health CAs were designed to provide informational support related to sexual and reproductive health topics for adolescents such as the definition of sex, birth control, testing for STI symptoms, and signs of pregnancy~\cite{bonnevie2021layla}. Along with accessibility, the major benefits of conversing with CAs for sexual health knowledge discovery include their ability to have non-judgemental conversations on confidential sexual health topics~\cite{park2020can}. Adolescents also perceived that interacting with sexual health CAs helped them reduce mental stress while navigating sensitive and often time, complex sexual health topics~\cite{rahman2021adolescentbot}. 

\section {Key challenges to address}
Broadly, there are two technical approaches to building CAs: LLM-based and Rule-based. We outline key challenges related to two approaches that we would like to address during the Child-Centered AI workshop. 

\paragraph {\textbf{Challenges in Rule-based CAs: Restrictive and Less-Human Like Responses}}
Rule-based CAs are developed with pre-defined keywords and commands programmed by the developer. Most of the CAs in the healthcare domain were built upon pre-defined sets of responses based on domain-specific knowledge. 
This means that the users are restricted to receiving predetermined answers to their questions, and there is little or no room for free responses. 
Early evidence showed that rule-based CAs were considered restricted in offering personalized advice, leading to low trust in the effectiveness of CAs in providing advice on mental and/or sexual health topics~\cite{nadarzynski2021barriers}. Particularly, rule-based sexual health CAs were seen as only providing advice about mainstream, easily accessible information, already available on the Internet. Subsequently, some struggled to understand the need for chatbots in sexual health~\cite{nadarzynski2021barriers}. Yet, the majority of the existing research ~\cite{nadarzynski2021barriers, lescano2022iterative, rahman2021adolescentbot, wang2022artificial, bonnevie2021layla} implemented rule-based approaches to develop CAs to provide mental and sexual health information for adolescents.

\paragraph{\textbf{Challenges in LLM-based CAs: Unregulated, Offensive, and Inappropriate Responses}}
LLM-based CAs that are trained on presumably the entirety of web data have the potential to tackle the above challenges, with the ability to understand input text written in human language in prompts and generate the responses~\cite{jo2023understanding}.
With such capability, LLM-based CAs have the potential to undertake more complex tasks that involve greater interaction and reasoning~\cite{tudor2020conversational} such as having interactive conversations related to sensitive topics. 
Yet, research on LLM-based CAs is still sparse and in the early stage (almost nonexistent) in the mental and sexual health context. 
In addition, there are limitations and challenges already documented in designing LLM-based CAs. Since LLMs have learned a vast amount of online text, there is a risk that the conversation flow can go beyond directions intended by the CAs designer~\cite{jo2023understanding}. Particularly, the risks inherent to LLM-based CAs can introduce adolescents to new types of risks such as being exposed to developmentally inappropriate and/or inaccurate content~\cite{Kelly2023}. 
With human-like and authoritative responses from LLM-based CAs, it may be difficult for adolescents to distinguish accurate information and fabricated answers~\cite{Rudy2023}. 
Hence, designing developmentally appropriate and accurate CAs for teens is pivotal for promoting their online safety. 

\section{Future Directions}
Overall, the safe evolution of CAs for adolescents needs to be discussed at the intersection of AI technology, child-centered design, and clinical support. Yet, very little work has been done to promote the digital safety of adolescents while interacting with AI-based conversational agents. Below are a few suggestions for future directions.

\begin{itemize}
    \item Research on LLM-based CAs in general is still at the beginning. New approaches (e.g., prompt tuning) are needed to refine the models to generate developmentally appropriate and accurate content for adolescents.
    \item More research efforts involving adolescents in the design of AI-based systems are needed to fulfill their needs on what kind of advice they prefer and what concerns they may have in diverse mental and sexual health contexts. 
    \item While the effectiveness of mental and sexual health CAs has been examined by trials, safety aspects of the CAs are under-explored. Therefore, more research efforts to evaluate the safety of systems (e.g., data security) and responses generated by the systems (e.g., accuracy, developmental appropriateness) are needed. 
\end{itemize}

\section{Conclusion}

Our research interests align very well with the purposes of the CCAI workshop to identify the major challenges and practical safeguards for generative AI systems. We anticipate learning more about organizers' and distinguished speakers' ground-breaking research to promote the child-centered development of AI-based systems. In addition, participating in CCAI would be extremely beneficial for us to have discourse on the safe evolution of generative AI in health contexts and gain valuable insights from organizers and participants, which could potentially lead to future collaboration opportunities. While we have identified some of the safety issues of AI-based CAs for adolescents, we hope that participating in the worship will help us address some of the remaining challenges and come up with design implications for safer generative AI systems for adolescents.

\section{About the Authors}
\textbf{Jinkyung Park} is a postdoctoral scholar in the Department of Computer Science at Vanderbilt University. Her research focuses on Human-Computer Interaction to promote online safety for vulnerable populations.
\newline
\textbf{Vivek Singh} is an associate professor in the School of Communication and Information at Rutgers University. He designs AI systems that are responsive to human values and needs.
\newline
\textbf{Pamela Wisniewski} is an associate professor in the Department of Computer Science at Vanderbilt University. Her work lies at the intersection of Human-Computer Interaction, Social Computing, and Privacy. Her expertise helps her empower end users and teach students to understand the value of user-centered design and evaluation.


\bibliographystyle{ACM-Reference-Format}
\bibliography{0_Main}

\end{document}